# All you can stream: Investigating the role of user behavior for greenhouse gas intensity of video streaming

Paul Suski
Division Sustainable Production and Consumption
Wuppertal Institut für Klima, Umwelt, Energie
Wuppertal, Germany
paul.suski@wupperinst.org

Johanna Pohl
Center for Technology and Society
Technische Universitaet Berlin
Berlin, Germany
pohl@ztg.tu-berlin.de

Vivian Frick
Center for Technology and Society
Technische Universitaet Berlin
Berlin, Germany
frick@ztg.tu-berlin.de

*Abstract*— **The information and communication technology sector reportedly has a relevant impact on the environment. Within this sector, video streaming has been identified as a major driver of $CO_2$-emissions. To make streaming more sustainable, environmentally relevant factors must be identified on both the user and the provider side. Hence, environmental assessments, like life cycle assessments (LCA), need to broaden their perspective from a mere technological to one that includes user decisions and behavior. However, quantitative data on user behavior (e.g. streaming duration, choice of end device and resolution) are often lacking or difficult to integrate in LCA. Additionally, identifying relevant determinants of user behavior, such as the design of streaming platforms or user motivations, may help to design streaming services that keep environmental impact at a passable level. In order to carry out assessments in such a way, interdisciplinary collaboration is necessary. Therefore, this exploratory study combined LCA with an online survey (N= 91, 7 consecutive days of assessment). Based on this dataset the use phase of online video streaming was modeled. Additionally, factors such as socio-demographic, motivational and contextual determinants were measured. Results show that $CO_2$-intensity of video streaming depends on several factors. It is shown that for climate intensity there is a factor 10 between choosing a smart TV and smartphone for video streaming. Furthermore, results show that some factors can be tackled from provider side to reduce overall energy demand at the user side; one of which is setting a low resolution as default.**

I. INTRODUCTION

The demand for videos streamed over the Internet is continuously increasing and so are the resulting electricity demand and the corresponding greenhouse gas (GHG) emissions. In 2009 annual global IP traffic was estimated to amount to 0,17 Zettabyte (ZB) with IP video traffic (Internet video, IP video on demand, video files exchanged through file sharing, video-streamed gaming, and video conferencing) being responsible for almost 35% of total data traffic [1]. In 2017, annual global IP traffic had already risen to 1.5 ZB, with IP video being responsible for 75% of overall data traffic. It is estimated that IP video traffic will further increase to 82% of then 4.75 ZB global IP traffic by 2022 [2]. In a recent study it was estimated that 60% of global data traffic can be attributed to online video and – further divided – to Video on Demand (20% of global data traffic), Tubes such as YouTube (13% of global data traffic), pornographic videos (17% of global data traffic) and online video from social networks and other webpages (11% of global data traffic) [3]. The case of online video is therefore highly relevant when it comes to investigating causes and development of increasing environmental relevance of digital services [4].

Recently, there have been discussions about how and whether the growth in data traffic (and especially in online video traffic) can be regulated [3], [5]–[8]. However, an estimation of the environmental impact of increasing online video use poses numerous challenges. Inventory data for the whole information and communication technology (ICT) sector are mostly estimates, not comprehensive and difficult to access [9], in particular when it comes to estimates regarding ICT infrastructure (network and data center) [10]–[14]. Available studies estimated that the total electricity consumption of the ICT sector was about 4 to 7% of global electricity demand by 2012 [15]–[17]. Forecasts assume an increase to 7 to 12% of global electricity demand at present with about half of the ICT sector's electricity consumption being attributable to network and data center use [18].

Transferring these estimations to global data volumes suggests that individual use of online video streaming is contributing substantially to global electricity demand and resulting environmental impacts. There are already a few studies on the relationship between online video consumption, data traffic and associated environmental impacts. Schien et al. estimated the energy demand of digital media, concluding that overall energy demand strongly depends on the type of end device and access network [14]. These findings are supported by Shehabi et al. [19], who recommend to increase the energy efficiency of user devices and networks in order to limit the increase in energy consumption from online video streaming. Morley et al. [7] studied the relationship of online video consumption, peak Internet traffic and national electricity demand and reveal clear parallels. Preist et al. [20] estimated the global GHG emissions associated with streaming YouTube videos for one year and identify a yearly GHG reduction potential of up to 6% by enabling audio-only streaming of music. Widdicks et al [21] applied a mixed methods approach, investigated streaming habits. They found that paid platforms and their 'all-you-can-eat' contracts evoke binge watching, and that new practices are occurring (e.g. 'multi-watching', separate content is streamed simultaneously over several devices by one or several users) in the household, which leads to increased data traffic. Hence, the following factors can be identified that increase online video traffic:

- Viewing duration of online video
  - mobile internet enabling the parallelization of activities such as watching online video while commuting or waiting
  - 'All-you-can-stream' flat rates
- Resolution of online video
- Bandwidth enabling high-bandwidth content such as Ultra-High-Definition (UHD), or 4K, video streaming

These factors are partly determined by the provider's or producer's choice (e.g. platform contracts, options and default settings for video resolutions), partly they are determined by the user's choice (e.g. streaming duration, choice of end device), and often both actors have an influence. In this paper, we aim to examine the role of the user's streaming behavior for online video traffic and associated GHG emissions. We identified the streaming duration, the resolution of the video and the choice of the device as relevant user behaviors for video streaming. Yet for streaming behavior to become more sustainable, insights on the relevance of these parameters can help define sustainable streaming behavior. Accordingly, three research questions (RQ) are formulated: RQ1: How do the three parameters (streaming duration, resolution, end device) impact overall GHG emission from online video streaming?

Additionally, as peak loads from online video traffic may lead to further extensions of bandwidth, and multi-watching is increasingly becoming an issue, it seems vital to gain more insight into streaming behavior habits concerning platforms, daytimes, weekdays as well as the setting in which streaming is taking place.

RQ2: When and how often do individuals stream on which platforms?

Finally, to find ways to reduce online video traffic, determinants of user behavior were examined. Behavior is determined by individual factors (e.g. norms for consumption reduction, digital literacy) and contextual factors (e.g. paid platform membership) [22], [23].

RQ3: What are determinants for environmentally relevant video streaming behavior (streaming duration, resolution and end device choice)?

## II. METHODOLOGICAL BACKGROUND: THE USER PERSPECTIVE IN ENVIRONMENTAL ASSESSMENT OF ICT-BASED SERVICES

Life cycle assessment (LCA) is one of the most established tools for assessing environmental effects of products or services. However, LCA studies often lack a proper scope or quality in the consideration of user perspectives, resulting in a high degree of uncertainty [4], [24]–[26]. As LCA intends to help decision-making to decrease environmental impacts, not only the descriptions and assessments of behavior is helpful. For example, also user-related factors can significantly alter environmental performance of ICT-based services [27]. Thus to support change, additionally contextual and user determinants can be identified that lead to specific behavior patterns and corresponding relevant environmental impacts [4]. For the case of online video streaming, we analyzed the relevance of user decisions and behavior and propose practical approaches that will allow for a better integration of user perspective in LCA studies (see *Figure 1*).

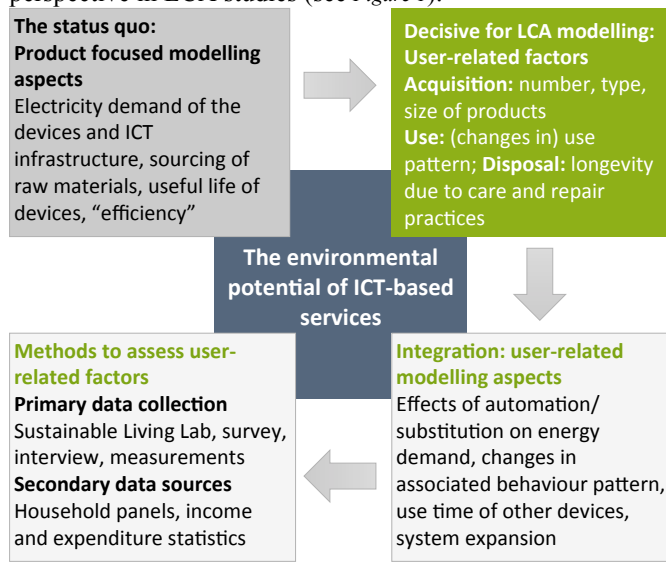

Figure 1: User decisions and behavior affecting the environmental potential of online video streaming

The inclusion of behavioral science is reported to be beneficial in LCA as a part of eco-design processes [24]. However, we argue that the benefits of a more realistic and holistic modeling and calculation is universal and not limited to eco-design. Increased knowledge of the specific utilization of products also allows a change in perspective, leading to more questions being answered by using different service units.

This approach shows that interdisciplinary methods or method combinations can achieve a more realistic modeling of user behavior in LCA. While LCA are often product focused, whether it be a device or a platform in the case of streaming, video-streaming users are put into focus here. As literature is lacking detailed data on video streaming behavior, primary data collection is essential. However, it has become clear from studies on sustainable consumption [28] that the responsibility for sustainable transformation does not lie solely with the consumer. Rather, the aim of consumer behavior analysis must be to strengthen the factors that promote sustainable behavior at both the supplier and political level and to eliminate factors that promote more environmentally intensive behavior.

### III. METHOD

In order to find answers to the research question, an online survey was conducted. Based on the survey results, an LCA focusing on video streaming behavior was performed. Resulting GHG emissions were then predicted by contextual and user determinants.

#### A. Sample

Participants were recruited as a convenience sample from the authors' social networks and are based in German speaking territories (Germany, Liechtenstein, Switzerland and Austria). Initially, $N = 121$ started filling out the survey on day 1. The final sample of $N = 91$. The dropout rate was 25 %. The initial and final sample did not differ significantly, so the drop-out was not biased. The sample was not representative: participants were younger, and had a higher education and income level, and there were more female participants than for example in the German population (*Table 1*) [29].

#### B. Survey design and procedure

Participants were invited to take part in a 7-day survey by sending them a link containing study information and terms of data handling, following General Data Protection Regulation. By entering their e-mail address, they consented to the data usage by the authors. The survey was done in Unipark, and there sent to the e-mail addresses on seven consecutive days in January 2020 (assessing Monday to Sunday). Each morning at 6 am, the survey link was distributed, and in the survey, the streaming behavior of the previous day was reported. On day 1, additionally socio-demographic variables were assessed, on day 7, covariates and determinants which are subsumed under measures, e.g. environmental concern, time spent online. At the end of the survey, participants could leave their comments in an open text field, they were thanked and informed that the debriefing would follow in summer 2020. Participants were allowed to fill in their data until three days after the survey week, as some had to be reminded and the goal was to minimize the dropout rate. The late debriefing time was chosen in order to prepare study results and advice for sustainable streaming behavior for participants as a way to thank them for their efforts.

Table 1: Sample of the online survey

| Age | *M(SD)* | 35.2 (10.7) |
|---|---|---|
| | *20-29 years* | 29.7 % |
| | *30-39 years* | 51.6 % |
| | *40-49 years* | 4.4 % |
| | *50-65 years* | 15.4 % |
| **Education level** | *primary* | - |
| | *secondary* | 7.7 % |
| | *tertiary* | 90.1 % |
| **Income level** | *< 1'500 €* | 33.0 % |
| | *1'500-3000€* | 43.9 % |
| | *> 3000 €* | 23.1 % |
| **Level of employment** | *full-time* | 34.1 % |
| | *half-time* | 39.6 % |
| | *unemployed* | 26.3 % |
| **Gender** | *female* | 61.5 % |
| | *male* | 37.4 % |
| | *other* | 1.1 % |

#### C. Measures

The measures are listed here in the order of appearance in the survey. We measured four aspects of streaming behavior: streaming duration, choice of device, audience size, and resolution of videos.

*Socio-demographic information*. Age, education level, income, level of employment and gender were reported.

*Membership at a paid platform*. For streaming subscriptions, participants indicated whether they had an account at Netflix, Amazon Prime Video, Apple TV+, Sky, DAZN or 'other'. Also, data packages of mobile data on the smartphone were assessed on the range of: no mobile data, 2 GB, 4 GB, 8 GB, to limitless data use, and the option 'I don't know'.

*Streaming duration*. Streaming duration was measured every day in four categories of video streaming: (1) on free video platforms (e.g. youtube, vimeo, youporn, illegal platforms), (2) paid platforms (e.g. Netflix, Amazon Prime Video, Apple TV+, DAZN), (3) on Social Media or on private websites (e.g. Facebook, Instagram, TikTok, Whatsapp), (4) in live streams or media centers from TV stations (e.g. Arte, ZDF, RTL, Zattoo). In this category, also watching regular broadcast TV was assessed (5). In each category, participants inserted the amount of hours they had watched videos on the day before during different time periods: (a) morning from 6:00 – 12:00, (b) afternoon between 12:00 - 18:00, (c) evening between 18:00 – 24:00 and (d) at night between 0:00 – 6:00 (d being assessed for the same day they filled out the survey). This 7-day assessment in the matrix based on (1-5) streaming platforms and (a-d) time periods, allowed for a differentiated picture of streaming durations over weekdays, daytimes and platforms.

*Streaming devices.* Each day, participants filled out which device they had mostly used when streaming, reporting the main device for each (1-4) streaming platform. They could choose between the devices laptop/PC, smartphone, smart TV and tablet, with the option of "does not apply to me" if they had not streamed on the platform in question. In consecutive calculations, the streaming device choice was integrated with streaming durations, based on each weekday, per platform.

*Audience size.* Equivalent to the assessment of streaming device, for each (1-4) streaming platform and (5) TV, on each day participants indicated whether they had watched alone, with one other person, two other persons, or more, and also had the option "does not apply to me" if they had not streamed on the platform in question.

*Video resolution.* At the end of the survey, participants indicated which resolution they had used most for each of the (1-4) streaming platforms. They could choose from: automatic, low (≤360p), middle (480p), HD (720p), ultra-HD (1080p).

*Parallel activities.* Each day, participants indicated which of the following activities they had enacted whilst streaming: housework (e.g. cooking, cleaning), mobility (e.g. in public transport), waiting time (e.g. for the bus or a doctors or business appointment), paid work, grooming (e.g. taking a bath), sports (e.g. treadmill), surfing the internet, other activities or no parallel activities.

*Unusual events.* The participants could indicate if anything unusual had happened in the week, which had inferred with their streaming behavior, choosing from sickness, holiday, stress, caretaking or adding other unusual events.

*Time spent online.* Participants estimated the number of hours they spend in their private time per day on fixed (e.g. PC) and mobile (e.g. smartphone) internet, each scale ranging from 0 to above 8 hours.

*Digital literacy and impact knowledge.* To measure digital literacy, we used a short, self-constructed 5-point Likert scale ranging from 'I know nothing about this' to 'I know a lot about this' including 3 items: general internet use (e.g. reading the news, shopping, sending mails, using social media), comprehension of technical processes (e.g. how the internet works, how to program a website), protecting one's privacy online (e.g. how to surf without being tracked, how to use an ad blocker). On the same scale, impact knowledge was measured with 1 item, knowledge on energy and resource use of digital appliances (e.g. streaming, cloud services).

*Environmental concern.* Environmental concern was measured by the 9-item scale developed in the German Environmental Concern Study (Umweltbewusstseinsstudie) [30].

*Personal norm for sufficiency-oriented streaming.* The personal norm [31] measuring the feeling of obligation to reduce personal video streaming impact was measured by two items on a 5-point Likert scale, e.g. 'Due to environmental reasons, I feel personally obliged not to stream too much.'

D. *Integration of Survey data in LCA*

We conducted a LCA based on ISO 14040 [32]. The aim was to assess the environmental effects related with online video streaming in order to determine factors for environmentally relevant video streaming behavior from the user side. The functional unit was online video streaming of one person during one week. Survey data were integrated into LCA by information regarding streaming duration per day, types of end devices and video resolution. For modeling the processes involved in the provision of the digital service 'video streaming', we followed the approach by [20] and included ICT infrastructure (access network, core and edge network, data center) and end devices into the system boundaries. The product system is depicted in *Figure 2*. We took a simplified approach regarding the number of life cycle stages and the number of impact categories taken into account and regarding the allocation of environmental impacts of products involved in the delivery of the service relative to data volumes. Furthermore, we did not collect any product-specific inventory data, but rely on recent data from the literature.

We provide results for global warming potential (GWP) over 100 years associated with online video streaming for one week. Except for end devices, we only included the environmental impact of service provision (electricity consumption during the use phase) in the modeling. As for the end devices, we also included the embodied emissions from production. The reason for this is that for carbon intensity of service provision embodied emissions for the production of end devices are of relevance, while they are negligible for ICT infrastructure [9]. In line with [20], our model does not take into account environmental effects associated with creation and upload of video content.

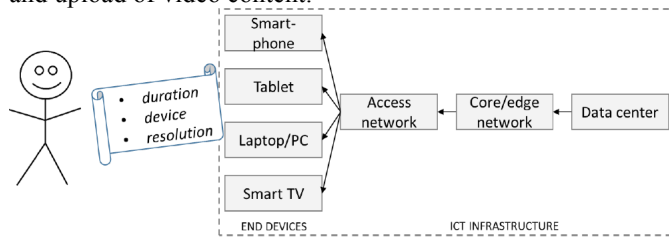

Figure 2: System boundaries for assessing online video streaming

In the following we describe for each part of the system, which data we have used and which further assumptions we have made.

*Digital Service*

As shown in *Table 2*, there are several devices and products involved in the delivery of online video. The provision of the service itself can be characterized by the resolution and the corresponding data traffic (bitrate). The actual resolution depends on many factors, which could not be considered in this study. Hence, assumptions were made based on regular native resolutions of the devices.

Table 2: Resolution and corresponding bitrate per streaming device

| Device | Standard resolution | Bitrate in GB/h |
|---|---|---|
| Smartphone | 360p | 0.3 |
| Tablet | 480p | 0.45 |
| Laptop / PC | 720p | 1.2 |
| Smart TV | 1080p | 1.8 |

*End device*

Both production and operation of the end devices are included in the modeling. For the inclusion of embodied emissions from production, we allocated embodied emissions proportionally to the share of streaming duration related to the daily overall use time of the device (based on statistical data for Germany for 2014 [33]). Electricity demand during online video streaming comes from the operation of the devices under medium load over the streaming duration. We assumed the German grid mix for assessing GWP. Important data and sources are presented in *Table 3*.

Table 3: Life cycle data for streaming devices

| Device | Embodied emissions in kg $CO_2$-eq. | Lifetime in years | Load in W | Source |
|---|---|---|---|---|
| Smartphone | 44 | 3 | 6 | [34], [35] |
| Tablet | 138 | 3 | 7 | [36], [37] |
| Laptop (PC) | 250 | 6 | 32 | [38] |
| Smart TV | 1000 | 8 | 200 | [39] |

*ICT infrastructure*

In order to take the environmental impact of data transmission of online video from the data center via core/edge and access network (the internet) to the home into account, a measure had to be found by which the environmental impacts of products involved in the delivery of the service can be determined proportionally. In line with [9], [11], [14], [40] we chose the energy intensity of transferring a specific volume of data in kWh/GB as measurement. As recommended by [11], we differentiated between data transmission via the core/edge and access network and data processing in data centers and tried to find the most up-to-date data possible. We assumed the EU-28 grid mix for assessing GWP [41]. Details are presented in *Table 4*.

Table 4: ICT infrastructure, which consists of access network, core and edge network and data center

| | Electricity intensity of data transmission in kWh/GB | Reference year | Source |
|---|---|---|---|
| **Access network** | 0.004 | 2014 | [42] |
| **Core and edge network** | 0.02 | 2014 | [43] |
| **Data center** | 0.049 | 2020 | [44] |

*Calculation of GHG intensity of streaming*

As described in the equation below, the GWP of online video streaming of one week results from the emissions due to production of the devices, the devices' operation and data transmission of online video.

$$GWP_{streaming} = GHG_{production} + GHG_{device\ operation} + GHG_{data\ traffic}$$
$$= \sum_{i=1}^{4} \left( D_i^{(W)} + \sum_{j=1}^{7} \sum_{k=1}^{4} t_{i,j,k}(P_i \gamma_G + R_i \rho \gamma_{EU}) \right)$$

| | | |
|---|---|---|
| $t_{i,j,k}$ | h | Streaming duration on device $i$ on day $j$ on streaming platform $k$ |
| $D_i^{(W)}$ | kg CO2-eq. | GHG from production per day for device $i$ |
| $P_i$ | kW | Power of device $i$ |
| $R_i$ | GB/h | Data traffic for resolution of device $i$ |
| $\rho$ | kWh/GB | Energy intensity of data traffic |
| $\gamma_G$ | kg CO2-eq./kWh | GHG intensity of electricity in Germany |
| $\gamma_{EU}$ | kg CO2-eq./kWh | GHG intensity of electricity in the European Union (EU28) |

IV. RESULTS

*A. Survey results on streaming behavior*

The survey showed that overall, participants had streamed $M(SD)$[1] = 1.77(1.22) hours per day. They had streamed more on weekends $M(SD)$ = 2.43(1.96) than on weekdays $M(SD)$ = 1.88(1.48), $t$ (90) = 3.79, $p < .001$, $D = .36$ As *Figure 3* illustrates, in the morning, $M(SD)$ = 0.24(0.29), afternoon $M(SD)$ = 0.44(0.52), and at night $M(SD)$ = 0.11(0.30), users streamed less than in the evening $M(SD)$ = 0.97(0.67).

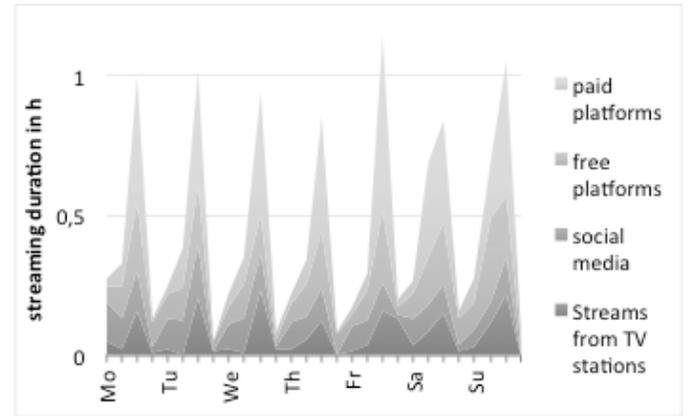

Figure 3: streaming activities over the week

People watched videos mostly on their laptops (almost 5 hours per week, *Table 5*), followed by smartphone and smart TV (3-4 hours), and only one hour on average was spent streaming on tablets. 72.5 % of participants had watched

---

[1] Statistical indicators:
$M$ = Mean; $SD$ = Standard deviation
$t$ = t value; $df$ = degrees of freedom; $D$ = effect size
$p$ = level of significance; + =.05; * = .01; ** = .001
$b$ = regression coefficient; $\beta$ = standardized regression coeff.
$r$ = Pearson correlation

videos on their smartphones, 65.9 % on a laptop/PC, 39.6 % on a smart TV and 20.9 % on a tablet.

Table 5: Weekly hours of streaming per device

|  | Laptop/PC | Smartphone | Smart TV | Tablet |
|---|---|---|---|---|
| M(SD) | 4.92(6.69) | 3.76(5.47) | 2.49(4.95) | 0.97(2.21) |
| Range | 0 - 40.5 | 0 - 25.5 | 0 - 25.5 | 0 - 17.0 |

Additionally, only between 4 – 15 % of the users had actively changed the resolution settings on platforms (*Table 6*), whereas the vast majority used automated settings. Presumably, participants answering they did not know which resolution they were streaming in, had not changed the default setting.

Table 6: Percentage of resolution choice per platform

|  | Free platform | Paid platform | Social Media | Streams from TV stations |
|---|---|---|---|---|
| automatic | 56.0 | 56.0 | 46.2 | 47.3 |
| low (≤360p) | 1.1 | 1.1 | - | - |
| middle (480p) | 3.3 | - | 2.2 | 3.3 |
| HD (720p) | 8.8 | 8.8 | 2.2 | 5.5 |
| u-HD (1080p) | 2.2 | 2.2 |  | 2.2 |
| I don't know | 28.6 | 31.9 | 49.5 | 41.8 |

*Table 7* shows the number of persons watching videos together. Videos on social media are mostly watched alone (1.06 persons on average). Watching television and video streams on paid platforms are the most social activities here with 1.63 and 1.58 persons watching together.

Table 7: Average number of persons watching a video

|  | N | M | SD |
|---|---|---|---|
| Free platform | 69 | 1.37 | 0.69 |
| Paid platform | 63 | 1.58 | 0.57 |
| Social Media | 49 | 1.06 | 0.32 |
| Streams from TV stations | 52 | 1.29 | 0.45 |
| Television | 35 | 1.63 | 0.66 |

*B. Environmental impacts of video streaming*

The environmental assessment of the video streaming behavior shows a high variety of the GWP, depending on streaming platform and devices (*Figure 4*). The GWP is displayed for the average streaming behavior over the whole week. The streaming duration is shown additionally on a second y-axis. Laptop and PC are the environmentally most relevant devices for watching videos on free platforms like YouTube (0,23kg $CO_2$-eq.), followed by the Smart TV (0.14 kg $CO_2$-eq.). Smartphone and Tablet are less relevant for watching videos on free platforms, even though the streaming duration on the smartphone is three times higher compared to the smart TV. This effect can be observed for all three subcategories: higher GWP for watching on smart TV due to device production, data traffic and electricity use to run the TV. For subscription-based paid platforms, the Smart TV is the most relevant device with resulting in 0.62 kg $CO_2$-eq., followed by the Laptop/PC with around half of the GWP. This is due to the high streaming duration on both devices and the high GWP effects for production, data traffic and electricity in the use phase. However, the higher streaming duration on Laptop and PC by 21 % compared to the smart TV still leads to a 46 % lower GWP due to lower resolution and production based emissions. The Smartphone shows barely an impact at all even though it was used around half an hour and the use of the tablet results in 0.1 kg $CO_2$-eq. Not surprisingly, videos on social media are mainly watched on smartphones (over 2 hours per week). This leads to the highest climate impact for this streaming category (0.07 kg $CO_2$-eq.). Laptop and PC play a minor role for social media videos with a streaming duration of 9 minutes per week on average and the Smart TV and Tablet show no environmental impact as they are barely used at all. The impacts from streams from TV stations are highest for the Laptop and PC with 0,18 kg $CO_2$-eq., which is in line with the highest streaming duration in this streaming category.

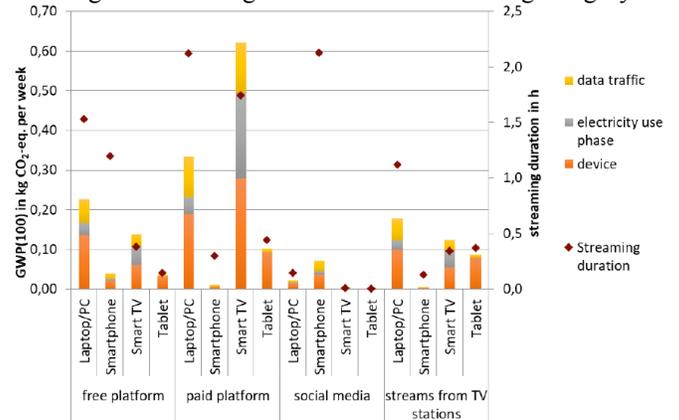

Figure 4: Global warming potential of average streaming activities and average streaming duration in one week

In *Figure 5*, the GWP is summarized for the four categories of streaming. It shows that watching videos on paid platforms result in the highest impacts (1.07 kg $CO_2$-eq.), which is in line with being the streaming category used the most with 4.6 h per week on average. Here, 22 % of the GWP are attributed to the data traffic. The electricity demand to run the devices contributes to 24 % of the GWP. The GWP from free platforms and streams from TV stations are on one level with 0.44 kg $CO_2$-eq. and 0.39 kg $CO_2$-eq. respectively. However, the streaming duration on free platforms is 65 % higher than the duration of streams from TV stations, indicating a use of more GWP-intensive devices for streams from TV stations. Here, we can observe in *Figure 4*, that free platforms are watched on low GWP intense Smartphones a lot, while streams from TV stations are watched a lot on tablets, which are of higher GWP intensity due to higher GHG emissions during production. Data traffic contributes to 24 % (free platform) and 22 % (paid platform). Video streams on social media platforms result in only 0.1 kg $CO_2$-eq. while being watched even more than streams from TV stations (2.3 h). This is due to being watched mainly on smartphones, resulting in low GWP in production, low data traffic due to low resolution and low electricity demand.

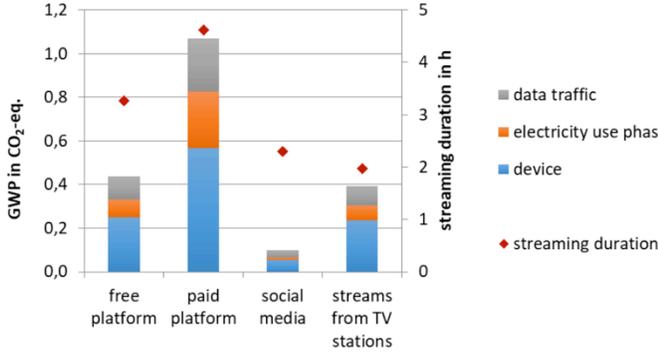

Figure 5: Global warming potential (GWP) of average streaming activities and average streaming duration in one week for four streaming categories

Figure 6 shows the GWP for video streaming sorted by choice of end device. The use of smart TV (0.89 kg $CO_2$-eq.) results in the highest GWP while being third regarding streaming duration (2.5 h). The smartphone shows opposite results with having the lowest GWP (0.13 kg CO2-eq.) and being used second most (3.8 h). Laptop and PC are being used most for streaming (4.9 h) and the data traffic contributes to 30% of the GWP. Data traffic is responsible for 21% of the GWP of the smart TV, while 35 % comes from the electricity demand in the use phase. Data traffic contributes to 8% of the GWP for the tablet and 36 % for the smartphone. The electricity demand in the use phase results in only 1% of the GWP of the tablet.

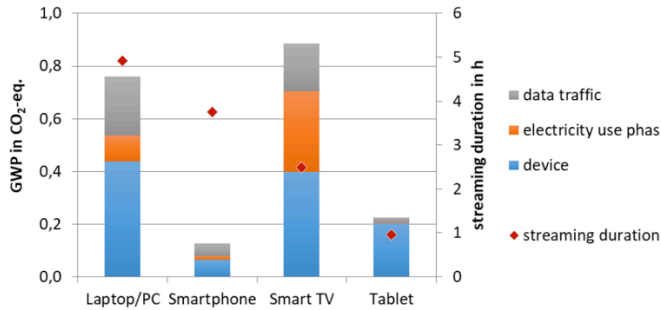

Figure 6: Global warming potential (GWP) of average streaming activities and average streaming duration in one week for four streaming devices

Now, the different GWP intensities of the device choices can be derived from the results (see *Table 8: GWP intensity for different devices per hour of streaming*. On average, the overall GWP for video streaming per person from this survey is 2 kg $CO_2$-eq. per week. Assuming that this represents average streaming behavior, it would add up to 104 kg CO2-eq. per year. The global $CO_2$ budget to stay below 1.5°C (with a 66 % chance) is 420 Gt $CO_2$ and the net emissions have to reach zero by 2050 [45]. This means, that until 2050 each person on the planet has, on average, a CO2-budget of 1609 kg annually (assuming an average of 8.7 billion inhabitants

[46]). This would imply that the video streaming behavior measured in our study takes up 6.5 % of a person's $CO_2$ budget.

Table 8: GWP intensity for different devices per hour of streaming

|  | Laptop/PC | Smartphone | Smart TV | Tablet |
|---|---|---|---|---|
|  | in kg CO2-eq. / h | | | |
| Device | 0.09 | 0.02 | 0.16 | 0.21 |
| Electricity use | 0.02 | 0.004 | 0.12 | 0.003 |
| Data traffic | 0.05 | 0.01 | 0.07 | 0.02 |
| Sum | 0.15 | 0.03 | 0.36 | 0.23 |

*C. Determinants of streaming behavior*

Multiple regression analyses were applied to identify predictors of daily streaming duration as a behavioral measure, and GHG emissions as an impact measure. Predictors in the models included socio-demographic (*Table 9* and *Table 10*) contextual and user-specific predictors of streaming behavior. As contextual predictors, we selected the membership at a paid platform (e.g. Netflix) and the size of the smartphone mobile data flat rate (ranging from 0, 2, 4, [...] GB to unrestricted flat rate, with a median of 2 GB). 77 % of participants had a paid membership on a streaming platform (compared to 56 % of people in Germany [47]). As user-specific predictors, we looked at the digital literacy, the knowledge about streaming impacts, and the personal norm for sufficiency-oriented streaming. Self-reported digital literacy was average overall, $M(SD) = 3.30 (0.73)$ on a scale from 1-5, compared to which impact knowledge was slightly lower, $M(SD) = 2.67 (1.19)$. The personal norm for sufficiency was also average, $M(SD) = 3.08 (1.13)$.

The membership at a paid platform significantly predicted both streaming hours and GHG emissions. The mobile internet flat rate size, as well as digital literacy and all socio-demographic predictors could not predict overall streaming duration and impact. Yet, the flat rate size of mobile internet could significantly predict the hours spent streaming on the smartphone in a simple regression analysis, $b(SD) = .23(.08)$, $\beta = .31$, $t = 2.94$, $p < .01$.

As for user-specific determinants, both the personal norm for sufficiency-oriented streaming behavior and the self-reported impact knowledge about streaming, although not significant in the regression models, were negatively correlated with daily streaming hours and GHG emissions. All these results and especially socio-demographic predictors have to be interpreted in context of the small sample size and the non-representatively distributed age and education level, which we will get back to in the discussion.

Table 9: Predictors of daily streaming duration, $R_2 = .17$

|  | b | SD | β | t | p | r |
|---|---|---|---|---|---|---|
| Digital literacy | 0.13 | 0.23 | .07 | 0.58 | .564 | .07 |
| Impact knowledge | -0.20 | 0.14 | -.18 | -1.39 | .169 | -.18[+] |
| Personal norm | -0.20 | 0.15 | -.17 | -1.30 | .198 | -.23[*] |
| Platform membership | 0.87 | 0.37 | .28[+] | 2.35 | .022 | .29[**] |
| Smartphone flat rate size | 0.00 | 0.02 | .00 | 0.02 | .986 | .00 |
| Age | -0.01 | 0.02 | -.05 | -0.34 | .733 | -.13 |
| Education level | -0.49 | 0.62 | -.10 | -0.79 | .435 | -.03 |
| Income | 0.00 | 0.07 | .00 | 0.00 | .998 | -.06 |
| Gender | 0.12 | 0.31 | .04 | 0.37 | .716 | .04 |

Table 10: Predictors of GHG emissions, $R_2$ equals .25

|  | b | SD | β | t | p | r |
|---|---|---|---|---|---|---|
| Digital literacy | 0.45 | 0.30 | .18 | 1.48 | .145 | .14 |
| Impact knowledge | -0.36 | 0.19 | -.23 | -1.86 | .068 | -.19[+] |
| Personal norm | -0.30 | 0.20 | -.18 | -1.49 | .142 | -.27[*] |
| Platform membership | 1.43 | 0.49 | .33[*] | 2.88 | .005 | .29[*] |
| Smartphone flat rate size | -0.04 | 0.03 | -.13 | -1.08 | .282 | -.15 |
| Age | 0.04 | 0.02 | .22 | 1.53 | .132 | -.03 |
| Education level | 0.63 | 0.83 | .09 | 0.76 | .452 | .13 |
| Income | -0.03 | 0.09 | -.04 | -0.27 | .786 | -.09 |
| Gender | 0.45 | 0.42 | .12 | 1.08 | .286 | .09 |

Further, we looked at parallel activities during streaming, which 94.5 % of users reported at least once. However, we find that only in a third of the reported streaming days, participants had engaged in parallel activities (*Table 11*).

Table 11: Percentage of streaming days at which participants reported parallel activities.

| Parallel activity | % of streaming days |
|---|---|
| No activities | 64.7% |
| Housework | 14.0% |
| Surfing the internet | 12.4% |
| Transfer time | 5.8% |
| Waiting time | 4.9% |
| Work time | 4.2% |
| Grooming | 3.3% |
| Sport | 3.1% |
| Other activities | 20.9% |

## V. DISCUSSION

Our study took an exploratory approach out the determinants of streaming behavior, exploring ways to integrate survey research from social sciences and psychology (behavioral data, behavioral determinants) with LCA methods.

We found that especially the choice of end device is decisive for overall online video streaming impact (RQ1), replicating the finding from Schien et al [29]. The smart TV shows the highest GWP intensity as a whole, as well as for each subcategory, the device production, the data traffic and the electricity use. On the opposite is the smartphone with a GWP intensity of around a tenth of the smart TV. In this context it is shown, that the resolution and the size of the display and the corresponding electricity demand are crucial factors for the GWP of online video streaming.

Looking at behavioral patterns (RQ2), we found that participants mainly streamed in the evenings and on weekends, and that the lion's share of streaming duration was spent on paid platforms. The choice of device however showed that laptops were used for the longest times, followed by smartphones and smart TVs, and the tablet was used the least. Still, the smartphone had been used by the most participants (about ¾), yet these watching episodes were overall shorter. Interestingly, only 15 % or less had changed default settings of video resolutions. Already in this rather well informed sample (recruiting in the field), 15 % or less of participants had changed resolution settings. Thus default settings have a strong impact on streaming impact. In the sample, parallel activities were done on a third of the reported streaming days. To understand streaming behavior and promote sustainable behavior, this parallelization must be further analyzed.

Regarding the determinants of streaming behavior (RQ3), we found that only the membership at a paid platform consistently predicted streaming hours and GWP. Yet these relationships may also be insignificant due to the small and non-representative sample. All in all, the results themselves have limited implications, as the convenience sample is not representative for the population: Participants were mainly highly educated, highly environmentally concerned and had higher internet use than the average population. Nevertheless, a few first conclusions can be drawn, which in further research may be backed up with more representative, and possibly long-term surveys.

The high relevance of online video streaming for achieving the goal of limiting climate change to 1.5°C was also underlined. 6.5 % of the available $CO_2$ budget is in opposition to the low amount of scientific publications, poor data availability and quality and overall public awareness.

Due shortages in data availability and sometimes poor data quality, several assumptions are included in the calculations of the GWP that need to be considered in the interpretation and for future research. A rather low resolution was assumed for smartphone and tablet, but technological developments that increase the native resolution of mobile devices and higher data packages for mobile data contracts are likely to increase the resolution and hence data traffic for such devices. Also, the calculations of the GWP did not consider mobile data traffic at all, leading to an underestimation of GWP for data traffic.

The electricity demand in the use phase is strongly influenced by the size of the display, leading to the conclusion that mobile devices will always show lower impacts due to electricity demands than stationary devices. Especially the increasing size of TV displays might have a relevant effect on future effects of video streaming.

As practical implications for environmental impact reduction for streaming, several possibilities may be offered. Recently, there have been discussions about how and whether the growth in data traffic (and especially in online video traffic) can be regulated [3], [5]–[8]. The problem of the high electricity demand of TVs can be addressed by changing the internal logic of the EU energy labels. Those indicate the energy efficiency of electrical devices. However, in the case of television devices, they compensate for display size, indicating the energy efficiency per inch of screen diagonal. As the size is the main driver, the energy label should only consider absolute energy demand and not relative demand to size.

The allocation of the climate impact of the production of the devices to online video streaming is another assumption that can be questioned, especially considering the high impact. Here, future studies might want to collect primary data on the usage of devices for online video streaming and other applications. The choice of the representative product for each device category is also open to discussion. We chose rather medium priced products for mobile devices and found only old data for the TV device. It is certain though that the production of the device is highly influential and hence so is its lifetime.

VI. CONCLUSION

The study shows, that online video streaming is environmentally highly relevant and depends on specific user behavior. By including behavioral science in environmental assessments we are able to understand variations in behavior to better address stakeholders to decrease environmental impacts from online video streaming. Streaming providers can adjust default settings for resolution, politics have to adjust the system of energy labels of TVs to not compensate for display size anymore. Additionally, there is a high demand in research on video streaming as technical and behavioral data is still missing. Representative surveys are necessary to further identify determinants for environmentally relevant online video streaming behavior.

ACKNOWLEDGMENT

Funding for this research was granted by the German Federal Ministry of Education and Research within the projects "Upscaling Strategies for an Urban Sharing Society" and "Digitalization and Sustainability" (www.nachhaltige-digitalisierung.de/en/). Every author contributed to this paper equally, independent from the place in the authors list.